\pdfoutput=1
\documentclass[11pt]{article}
\usepackage{geometry} 
\usepackage{graphicx} 
\usepackage[font=footnotesize,labelfont=bf]{caption}
\usepackage{float} 
\usepackage{wrapfig} 
\usepackage{amsmath}
\usepackage{cite}
\usepackage{csquotes}
\usepackage{color}
\usepackage[normalem]{ulem} 
\newcommand{\e}{\epsilon}
\newcommand{\w}{\omega}

\begin{document}
\title{Enhancement of Resolution and Propagation Length by Sources with Temporal Decay in Plasmonic Devices}

\author{H. Serhat Tetikol and M. I. Aksun\thanks{iaksun@ku.edu.tr (corresponding author)} \\Electrical \& Electronics Engineering, Ko\c{c} University, Istanbul, Turkey}

\maketitle

\begin{abstract}
Highly lossy nature of metals has severely limited the scope of practical applications of plasmonics. The conventional approach to circumvent this limitation has been to search for new materials with more favorable dielectric properties (e.g., reduced loss), or to incorporate gain media to overcome the inherent loss. In this study, however, we turn our attention to the source and show that, by imposing temporal decay on the excitation, SPP modes with simultaneous complex frequencies and complex wave vectors can be excited with enhanced resolution and propagation length.  Therefore, to understand the underlying physics of these phenomena and, in turn, to be able to tune them for specific applications, we propose a framework of pseudo-monochromatic modes that are generated by introducing exponential decays into otherwise monochromatic sources. Within this framework, the dispersion relation of complex SPPs is re-evaluated and cast to be a surface rather than a curve, depicting all possible $\w-k$ pairs (both complex in general) that are supported by the given geometry. Since the improvement in resolution and propagation length due to the introduction of temporal decay to the excitation is rather counter-intuitive (i.e., adding temporal loss improves the propagation length), the dispersion-based theoretical predictions have been validated via the FDTD simulations of Maxwell's equations in the same geometry without any a priori assumptions on the frequency or the wave vector. Moreover, improvement in resolution with the temporal decay has been demonstrated in a plasmonic superlens structure to further validate the predictions.



\end{abstract}


\section{Introduction}
Surface plasmon polariton (SPP) is a coupled electromagnetic and electron wave that propagates along the interface between two dissimilar media, usually dielectric and metal.  This harmonious union of electromagnetic wave and electrons, i.e., collective oscillations of surface charge density, is confined to the vicinity of the interface while propagating a rather long distance along the interface, as compared to the SPP wavelength. Therefore, SPPs provide a unique set of features including highly localized energy and enhanced field strength near the interface, rendering it very sensitive to surface structures. As such, SPPs have been quite instrumental to improve devices such as sensors, waveguides, photovoltaic cells, optical microscopes, photodetectors and modulators \cite{barnes2003surface,ebbesen2008surface, schuller2010plasmonics,atwater2010plasmonics}. Furthermore, SPPs have been proven to be the underlying cause of phenomena such as extraordinary transmission, negative refraction, superlens, and metasurfaces.

Although there are many such applications that SPPs can either facilitate or improve, their practical realizations have been severely limited by high intrinsic loss that naturally exists in metals \cite{stockman2018roadmap,naik2013alternative}. To remedy this inherent shortcoming of SPPs, an extensive search for materials with more favorable dielectric properties has already been conducted and is still on-going \cite{west2010searching,naik2013alternative,boltasseva2011low,stockman2018roadmap}. In this paper, however, we shift our attention towards the time signature of the excitation source and study the response of layered plasmonic structures to exponentially decaying but otherwise monochromatic excitations (complex frequency), with a view to mitigate the stated shortcoming.  Since the SPPs excited by such sources can be represented by the eigenfunction of  $e^{j\w t - jkx}$, with associated complex eigenvalue pairs ($\w$, $k$), we can still use the frequency-domain approach and obtain the dispersion relation analytically, as it has been the case for monochromatic excitations. The use of a signal with complex frequency as the source does not only lend itself for a well-known and easy-to-use frequency-domain analysis, but also provides a convenient framework to study the effects of a spatio-temporal source and design its time signature for an intended application. Within this framework, the dispersion equation has been re-examined and visualized over the entire complex domains of $\w$ and $k$ as a \textit{surface}, instead of a curve. As a result, the range of possibilities for dispersion engineering has been dramatically broadened through the excitation of modes on the dispersion surface, allowing one to control and tailor the salient features of SPPs.


Several recent studies have introduced temporal variations to the excitation source for the SPPs to obtain improved resolution in flat imaging devices \cite{rogov2018space, dubois2015time, archambault2012superlens}. Although these studies have played an important role to demonstrate the importance of time signature of the excitation, they have been rather limited in scope and application mainly because the time variations of the source that they have used are either convoluted or not easy-to-generalize, and more importantly, they do not provide a framework to study different time signatures for different purposes. Therefore, in this study, we aim to generalize the use of non-monochromatic time varying sources for the excitation of SPPs and to propose a framework that will enable us to select a dispersion curve out of a dispersion surface for better and improved performance criteria, like resolution, propagation length and field enhancement.

\section{Results}
\subsection{Dispersion surface for SPPs in metal-insulator-metal waveguides}
\begin{figure}[htbp]
\centerline{\includegraphics[width=10cm]{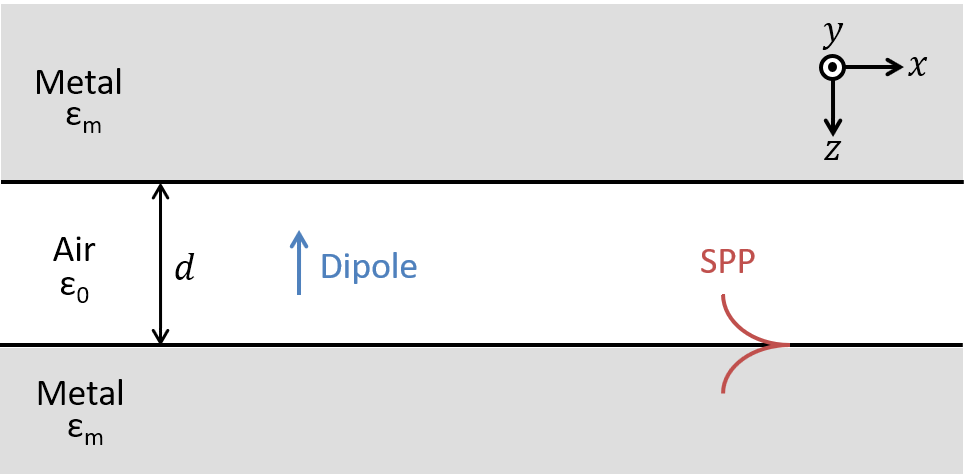}}
\caption{A typical metal-insulator-metal plasmonic waveguide. A vertical dipole, placed in the insulator of thickness $d$, is used to excite the SPP modes of the entire wave vector spectrum. At sufficiently large distances away from the dipole along the interface, only the SPP modes would survive as they would be the only modes supported by the waveguide within the frequency range of interest.}
\label{MIM_waveguide_dipole}
\end{figure}

Although the approach proposed in this work is quite general for planar plasmonic waveguides, for the sake of illustration and with no loss of generality, we have chosen a metal-air-metal structure with two semi-infinite metal layers of the same material and an air gap of $50nm$ in between, as shown in Figure \ref{MIM_waveguide_dipole}. Since the SPP modes of layered geometries can be obtained simply by finding the poles of the generalized reflection (or the transmission coefficient) \cite{chew1995waves,maier2007plasmonics}, the expression of the generalized reflection coefficient at the air-metal interface is given as
\begin{equation}
	\tilde{r} = \frac{-r+re^{-jk_z2d}} {1-r^2e^{-jk_z2d}}
\label{eq:genr}
\end{equation}
where $r$ is the Fresnel reflection coefficient at the air-metal interface, $k_z$ is the normal component of the wave vector in free space, and $d$ is the separation between the two semi-infinite metal layers. Note that, in this study, the relative permittivity of the metal layers $\e_m$ is approximated by the Drude model~(\ref{eq:drudeModel}), which agrees well up to ultraviolet or visible frequencies with the experimental data for metals such as silver, gold, copper and aluminum \cite{johnson1972optical,babar2015optical,mcpeak2015plasmonic},
\begin{equation}
	\e_m = \e_{\infty} - \frac{\w_p^2}{\w^2-j\gamma\w}
\label{eq:drudeModel}
\end{equation}
where $\w_p = 1.04\times 10^{16}\ rad/s$ is the plasma frequency of the metal, $\gamma = 6.15 \times 10^{14}\ rad/s = 0.059 \times \w_p$ accounts for the loss in the metal, and $\e_\infty = 4$ is the dielectric constant at very high frequencies ($\w\gg\w_p$).

In finding the electromagnetic modes of a given structure, the conventional approach has been to find all possible solutions of the associated dispersion relation for either real frequencies or real wave vectors. These two choices result in non-identical dispersion relations with fundamental differences for the same geometry; an observation that was made in the field of plasmonics decades ago \cite{arakawa1973effect,alexander1974dispersion,kovener1976surface}. Similar observations have been made in linear chains of metallic nanospheres \cite{koenderink2006complex,conforti2010dispersive,udagedara2011complex}, cylindrical metallic nanowires \cite{wan2012analytical}, layered media \cite{archambault2009surface}, and grating structures \cite{weber1986determination,barnes1996physical}. Indeed, this is not limited to SPPs, and the same situation emerges in photonic crystals \cite{huang2004nature}, bulk polaritons \cite{barker1972response}, acoustic materials \cite{fang2006ultrasonic} and virtually any real physical problem that gives rise to a dispersion relation. Other discussions on this topic are available in \cite{le2008principles,archambault2009surface}.
However, we have to note that, in a few recent studies \cite{archambault2012superlens,baranov2017coherent,dubois2015time, rogov2018space}, time-domain signals with non-monochromatic time signatures have been used and shown to improve resolution and enable access to electromagnetic modes otherwise not possible. In this study, our aim has been to provide the basis of these observations, and in turn, to propose a general tool to help select the most suitable dispersion curve for a given application. This has been achieved by allowing both frequency and wave vector to take complex values simultaneously, with the imaginary parts representing temporal and spatial damping, respectively. As a result, a dispersion surface comprising complex SPP modes emerges instead of a simple dispersion curve corresponding to either real frequencies or real wave vectors. Consequently, it has been shown that, due to these complex SPP modes, substantially higher resolution and longer propagation lengths can be realized from the same geometry and materials.

\begin{figure}[htbp]
\centerline{\includegraphics[width=15cm]{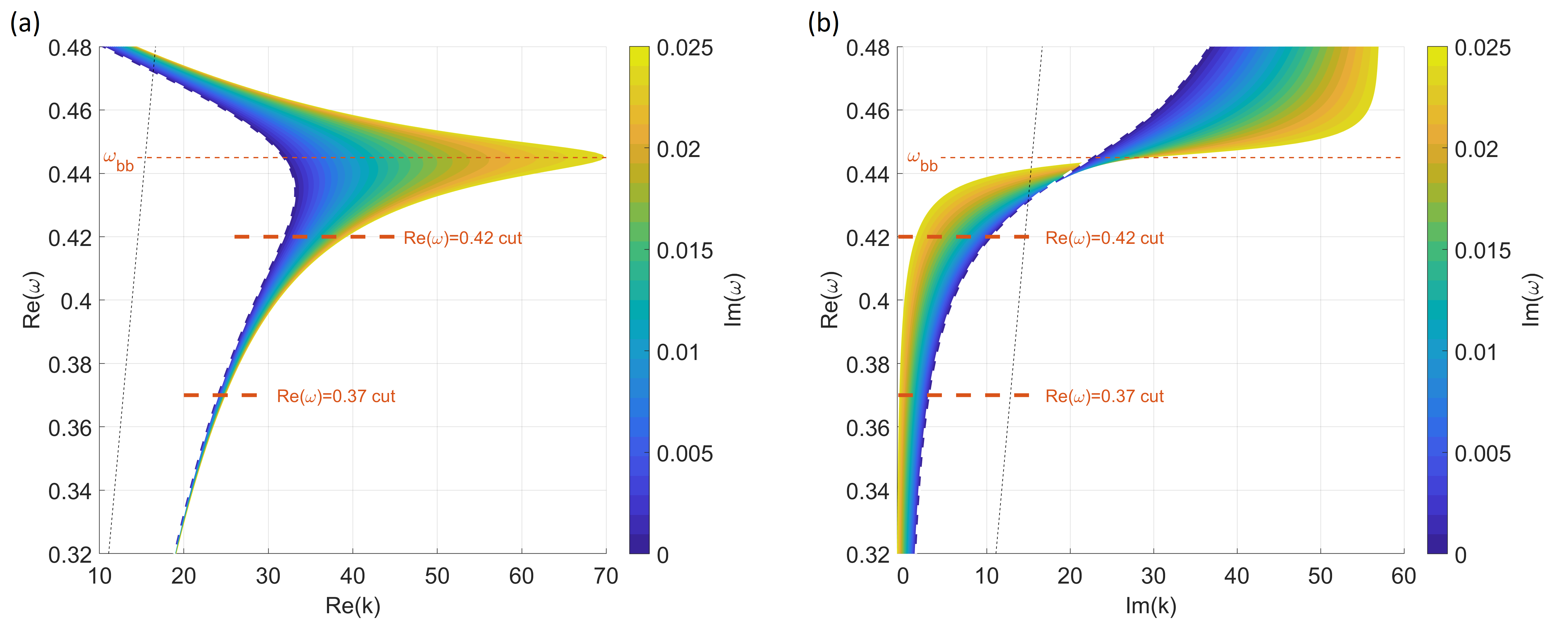}}
\caption{The dispersion surface for the metal-air-metal waveguide shown in Figure \ref{MIM_waveguide_dipole}. \textbf{(a)} $Re(\w)$ vs. $Re(k)$, and \textbf{(b)} $Re(\w)$ vs. $Im(k)$ for a range of temporal decay rates $Im(\w)$ represented by the color axis. The dark blue edge of the surface (delineated by a dashed line) corresponds to the case with zero temporal decay, i.e., $Im(\w)$=0. Please note that, for frequencies below the back-bending frequency $\w_{bb}$, increasing the temporal decay (towards lighter color in both) leads to larger wave vectors with smaller imaginary part. This is the key finding of this study as it provides an opportunity to improve the resolution and the propagation length of SPPs, which will be elucidated further by investigating the modes along the $Re(\w)=0.42$ and $Re(\w)=0.37$ cuts on the surface (red dashed lines). For the experimental configuration in Figure \ref{MIM_waveguide_dipole}, the temporal decay of the SPP modes can be independently controlled by driving the dipole with the desired time signal. The dashed black line is the light cone in free space.}
\label{MIM_disp_surface_combined}
\end{figure}

All $\w,k$ pairs that satisfy the dispersion relation are calculated and displayed as two separate surfaces in Figures \ref{MIM_disp_surface_combined}a and \ref{MIM_disp_surface_combined}b for the real and imaginary parts of the wave vector, respectively. The color axis represents the imaginary part of the frequency, i.e., the temporal decay rate. For the sake of providing perspective, the conventional dispersion curve, for which real frequencies and complex wave vectors are assumed a priori, is shown by the dashed line at the edge of the dark blue region in Figure \ref{MIM_disp_surface_combined}.

We identify two main regions on the dispersion surface: $Re(\w)>\w_{bb}$ and $Re(\w)<\w_{bb}$, where $\w_{bb}$ marks the frequency corresponding to the tip of the back-bending region. Since, in $Re(\w)>\w_{bb}$ region, the imaginary part of the wave vector (Figure \ref{MIM_disp_surface_combined}b) increases to prohibitively large values for any practical applications, we focus only on the modes below the back-bending frequency $\w_{bb}$, where the contributions of this work, namely, the enhancements of resolution and propagation length, are more pronounced and visible. The key observation is, with increasing temporal decay ($Im\{\w\}$), indicated by the lighter colors on the dispersion surface, the real part of the wave vector ($Re\{k\}$) reaches much larger values compared to the conventional dispersion curve with real frequencies (Figure \ref{MIM_disp_surface_combined}a) while the imaginary part of the wave vector moves closer to zero (Figure \ref{MIM_disp_surface_combined}b), which are the clear manifestations of better resolution and longer propagation length of the corresponding SPPs, respectively.

The main conclusion that has been derived from the study of dispersion relation in this section can be stated as follows: {\it "sources decaying exponentially in time may excite complex SPPs that have higher resolution and longer propagation length"}. However, there are two legitimate concerns that need to be addressed, namely 1) if it is valid to use complex $\w$ and $k$ in the formation of the dispersion surface, rather than the conventional choice of real $\w$ and complex $k$, and 2) if it is possible to validate longer propagation length for the sources with decaying time signature, as being rather counterintuitive. Moreover, calculation of the propagation length when a temporal decay is introduced to the input signal needs to be re-defined as it should involve both temporal and spatial attenuation; that is, the propagation length can not be as simple as the reciprocal of the imaginary part of the wavevector anymore. Although the first concern can be resolved mathematically by showing that the eigenvalues of the wave equation in open structures are in general complex, both concerns can be alleviated at once by simply solving the same geometry by using a full-wave EM simulator, like those based on Finite Element Method or Finite-Difference Time-Domain method, without any prior assumption on the frequency or the wave vector.

\subsection{SPP resolution, propagation length and lifetime}

In this section, we investigate, in detail, the salient features of the SPPs with temporal decay by using the FDTD simulations of Maxwell's equations, and compare and validate the theoretical results, i.e., the dispersion surface (Figure \ref{MIM_disp_surface_combined}). For the FDTD simulations of the waveguide in Figure \ref{MIM_waveguide_dipole}, we use Lumerical FDTD solutions \cite{lumerical} and observe the fields starting at a sufficiently large distance away from the dipole in order for the SPPs to be the only remaining wave in the waveguide. Once all the field components are collected over both time and space along the propagation direction, we use a numerical method, based on the generalized pencil of function (GPOF) method, to extract the frequency and the wave vector of the propagating SPP modes (see Methods) \cite{hua1989generalized}.

For the sake of elucidating the perceived improvements in the features of the SPPs upon studying the dispersion surface, we have chosen two samples of $Re(\w)$: $Re(\w)=0.42$ for higher frequencies closer to $\w_{bb}$ and $Re(\w)=0.37$ for lower frequencies away from $\w_{bb}$, as delineated in Figure \ref{MIM_disp_surface_combined}. The characteristics of the modes at both frequencies are analyzed in detail, both theoretically and using the FDTD simulations, and  their improvements upon introducing temporal loss into the system have been demonstrated.

\begin{figure}[htbp]
\centerline{\includegraphics[width=16cm]{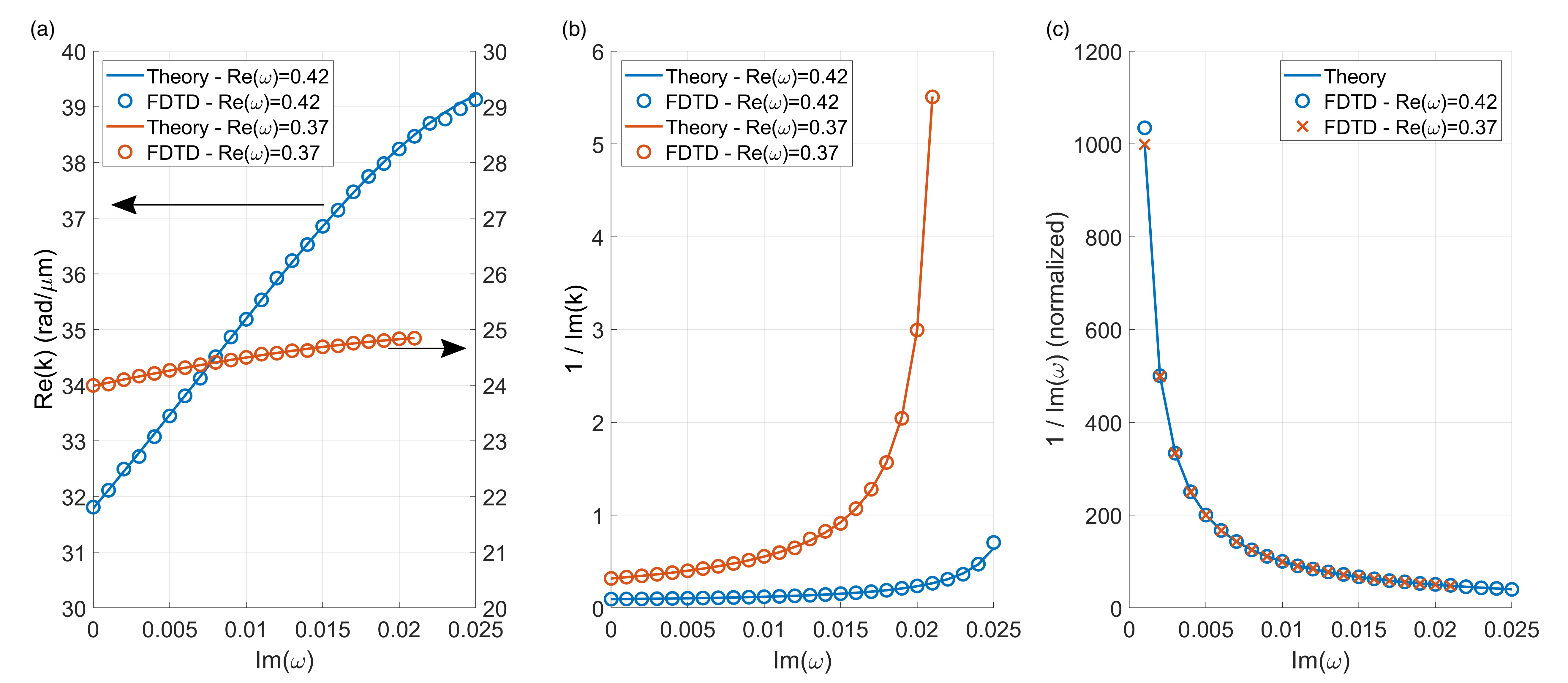}}
\caption{Salient characteristics of the SPPs excited in the metal-air-metal waveguide (Figure \ref{MIM_waveguide_dipole}) for the frequencies of $Re(\w)=0.42$ and $Re(\w)=0.37$. An excellent agreement between the theory and the FDTD simulations is observed for all metrics. \textbf{(a)} The plot of $Re(k)$ vs. $Im(\w)$ demonstrates a significant improvement in resolution, especially at $Re(\w)=0.42$, with the increase of temporal decay. The left and right y-axes are for the $Re(\w)=0.42$ and $Re(\w)=0.37$ curves, respectively. \textbf{(b)} The plot of $1/Im(k)$ vs. $Im(\w)$ demonstrates an increase in the propagation length of the SPP. Especially for $Re(\w)=0.42$, it is more than an order of magnitude compared to the time harmonic case ($Im(\w)=0$). However, note that $1/Im(k)$ loosely represents the propagation length for non time-harmonic cases, for which the effective propagation length is shown in Figure \ref{MIM_effective_propagation_length}. \textbf{(c)} The reciprocal of the imaginary part of the frequency indicates the lifetime of the SPP. Since the temporal decay is independently controlled by the dipole excitation source, it is exactly the same for both $Re(\w)=0.42$ and $Re(\w)=0.37$ curves.}
\label{MIM_resolution_curve}
\end{figure}

The resolution of an SPP is directly related to the real part of its wave vector along the propagation direction, that is, if the wave vector is large enough that the wavelength of the SPP is comparable or smaller than the dimensions of an irregularity near or at the interface, the SPP fields scatter in measurable amounts and can be detected with a proper experimental setup. For the modes of interest, namely those corresponding to $Re(\w)=0.37$ and $Re(\w)=0.42$, the real part of the wave vector vs. temporal loss $Im(\w)$ is given in Figure \ref{MIM_resolution_curve}(a), where the agreement between the theory and the FDTD simulations is excellent. For both frequencies, the resolution increases with increasing temporal decay, as expected. While it is relatively small for modes corresponding to $Re(\w)=0.37$, we see more than 20\% improvement in the resolution for the modes with $Re(\w)=0.42$. As also seen in the dispersion surface, the theoretical upper limit on resolution, which can be attained by introducing temporal decay to the excitation, grows with the real part of the frequency. For the range of frequencies shown in the dispersion surface, it is possible to more than double the resolution near the back-bending region.

Another feature of the SPPs that is crucial for applications is its lifetime, which is proportional to the reciprocal of the imaginary part of the frequency. Since we have used the same temporal decay rates at both frequencies, $Re(\w)=0.42$ and $Re(\w)=0.37$, their lifetimes obtained from the simulations are expected to coincide with each other, and moreover, to be in good agreement with the theory, both of which are demonstrated in Figure \ref{MIM_resolution_curve}(c). As observed, introduction of even a very small amount of temporal decay to the time-harmonic excitation results in a dramatic decrease in the lifetime of the SPP, while further increases in temporal decay have a smaller effect on the lifetime.

Since the reciprocal of the imaginary part of the wave vector, $1/Im(k)$, is usually associated with the propagation length, as the key part of this study, it is given in Figure \ref{MIM_resolution_curve}(b), demonstrating an excellent agreement  between the theory and the simulations. It is important to note that $Im(k)$ decreases with increasing temporal decay, and it appears as if the propagation length can be increased by an order of magnitude. However, the definition of $1/Im(k)$ (or $1/Im(2k)$ to be precise) as the propagation length is only valid for the time-harmonic excitations where the attenuation of the wave is attributed solely to the spatial decay. For the SPP modes with complex frequencies, which experience \emph{temporal decay} in addition to the spatial attenuation, one must take into account the \emph{propagation time}. Therefore, defining an effective propagation length $l_{eff}$ is in order, as we would like to compare our results against the time-harmonic case. Since an SPP mode on the dispersion surface (Figure \ref{MIM_disp_surface_combined}) has indeed a single complex frequency, we can write it in the form of $e^{j\w t-jkx} = e^{j\w' t-jk'x}e^{-\w'' t-k''x}$ where single and double primes represent real and imaginary parts, respectively. Hence, one can define the effective spatial decay parameter $\alpha$ by simply equating the amplitude of this expression to $e^{-\alpha x}$, resulting in
\begin{equation}
	l_{eff} = \frac{1}{\alpha} = \frac{v_{spp}}{\w'' + k''v_{spp}} = \frac{\w'}{\w''k'+\w'k''}
\label{eq:effectivePropagation}
\end{equation}
where $v_{spp}$ is the propagation velocity of the SPP along the interface, and can be calculated directly from the dispersion relation.

\begin{figure}[htbp]
\centerline{\includegraphics[width=12cm]{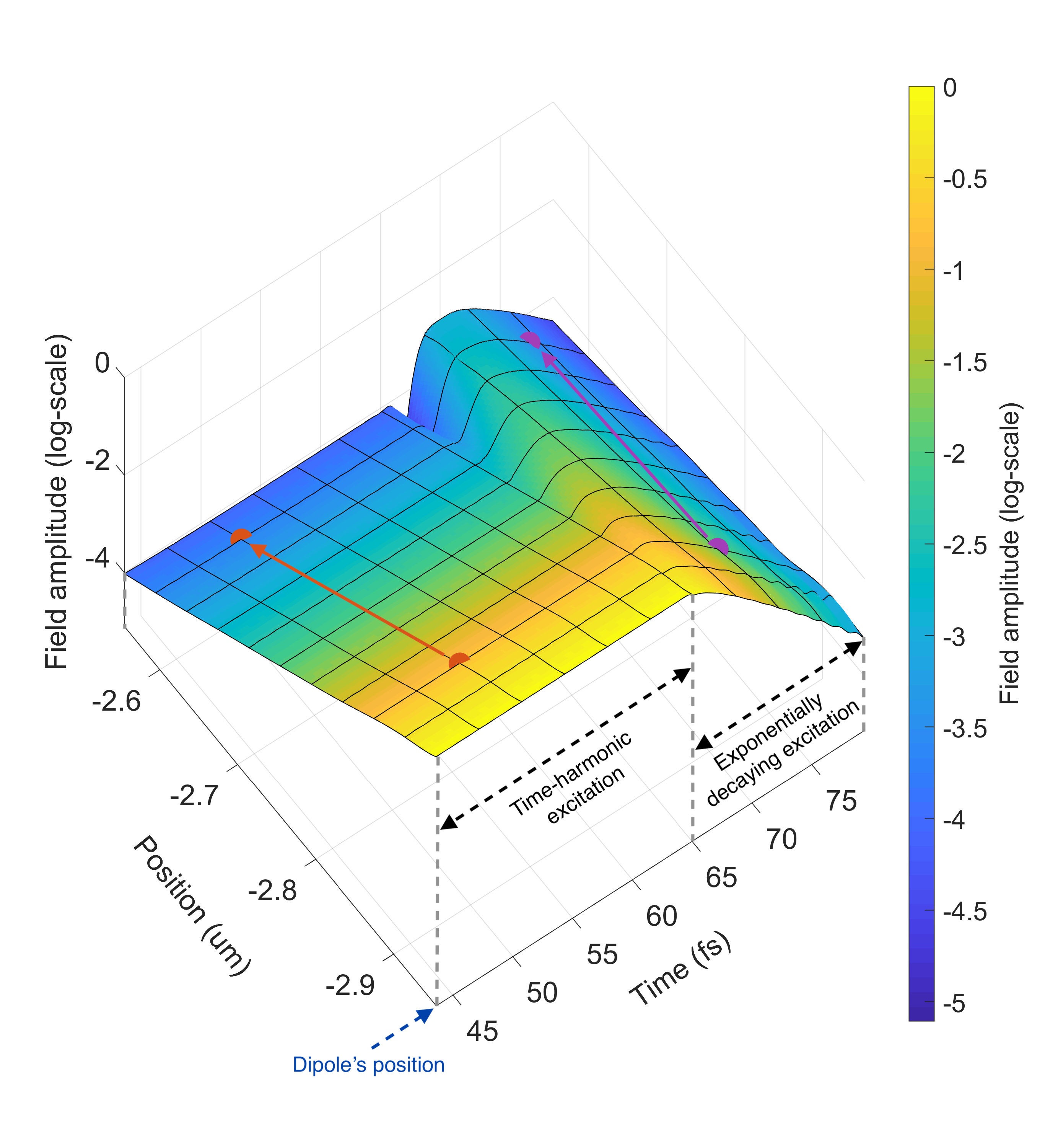}}
\caption{Calculation of the effective propagation length of the SPP from the FDTD simulation results. The vertical electric dipole, positioned within the plasmonic waveguide as shown in Figure \ref{MIM_waveguide_dipole}, first excites the waveguide with a time-harmonic signal ($\w=0.42+0j$). After the system reaches the steady-state, the temporal decay is introduced and the SPP with a complex frequency of $0.42+0.025j$ begins to propagate within the waveguide. A point is selected on the SPP's field profile and its amplitude is observed after some time, as shown with the purple dots and arrow. The ratio of the initial and the final amplitudes can be used to calculate the effective propagation length $l_{eff}$. For comparison, we have also included the equivalent path for the steady-state case, shown in red, which coincides with the constant-time lines of the black grid, while the complex frequency path is tilted. This is because one needs to take both time and space into account for temporally varying field profiles, whereas time does not matter for time-harmonic waves as the field is constant for all times at a given position. All measurements are made long after the switching from time-harmonic to complex frequency excitation.}
\label{MIM_point_tracking}
\end{figure}

For the sake of validating the theoretical effective propagation length given in (\ref{eq:effectivePropagation}) and the conclusions drawn from it, we obtain the propagation length from the FDTD simulations as follows: i) calculate the envelope of the SPP fields in time domain, and ii) track a selected point on the envelope as it propagates along the waveguide. To elucidate, an example is provided in Figure \ref{MIM_point_tracking}, where the waveguide is first excited by a time-harmonic excitation, and upon reaching its steady-state, the temporal decay is introduced on the excitation that excites the SPP with the complex frequency of $0.42+0.025j$. Then, we select a point in the exponentially decaying region (shown in purple) and measure how much it decays as it propagates, from which one can easily deduce the effective propagation length from the simulation.

\begin{figure}[htbp]
\centerline{\includegraphics[width=10cm]{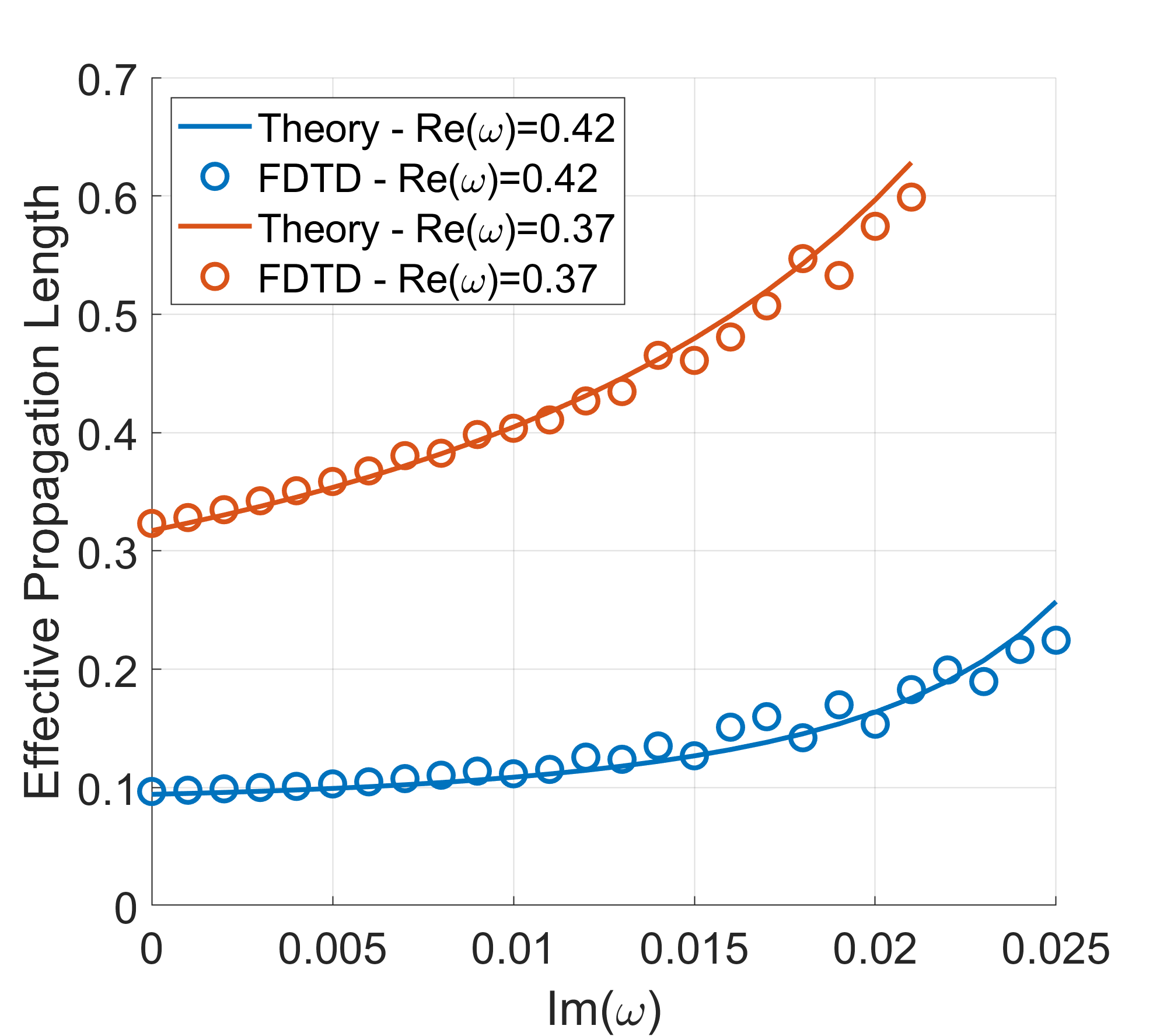}}
\caption{Effective propagation lengths vs. temporal loss for the modes along the $Re(\w)=0.42$ and $Re(\w)=0.37$ cuts on the dispersion surface (Figure \ref{MIM_disp_surface_combined}). The theoretical curves, obtained from equation (\ref{eq:effectivePropagation}), and the FDTD results, calculated as explained in Figure \ref{MIM_point_tracking}, are in good agreement for all values of $Im(\w)$. Increasing the temporal decay results in longer effective propagation length for both frequencies while the lower frequency SPPs ($Re(\w)=0.37$) propagate much further than the higher frequency SPPs ($Re(\w)=0.42$). Further improvement is possible with faster temporal decay rates. The slightly larger variations at higher $Im(\w)$ values are due to the numerical noise caused by rapidly-changing amplitudes in the FDTD simulations.}
\label{MIM_effective_propagation_length}
\end{figure}

As the effective propagation length has been well-defined in cases of non-monochromatic excitations, both in theory and in the FDTD simulations, their comparisons provide additional validation for the assessment of longer propagation length when temporal loss is introduced, as shown in Figure \ref{MIM_effective_propagation_length}. It is observed that the effective propagation length, $l_{eff}$, is approximately doubled for both frequencies ($Re(\w)=0.42$ and $Re(\w)=0.37$) over the range of $Im(\w)$ analyzed in this study, and it can lead to orders of magnitude increase in the field enhancement depending on the distance between the excitation and the observation points. Further improvement is also possible with larger temporal decay rates; however, it becomes increasingly difficult to simulate rapidly-changing amplitudes by the FDTD method. Notice that the doubling of the effective propagation length is in contrast with the ten-fold improvement one would expect from Figure \ref{MIM_resolution_curve}(b). Since the finite lifetime of SPPs partially counteracts this improvement, it is critical to take both the wave vector and the frequency into account in order to study the propagation characteristics of complex SPP modes. As an illustration, a detailed propagation movie of a complex SPP mode selected from Figure \ref{MIM_effective_propagation_length} is provided in the Supplementary Movie.

\subsection{Superlens: Application of dispersion surface}
Superlens structures are, in general, made out of metamaterials with negative permittivity and permeability, and provide resolving power beyond the conventional diffraction limit of $\lambda/2$. Moreover, plasmonic structures, as simple as a thin metal film, can also display superlens properties in the nearfield where the electrostatic approximation holds \cite{pendry2000negative}.
In this section, we analyze a Superlens structure of the latter type within the pseudo-monochromatic framework and show that its resolving power can be improved using the dispersion surface concept. For the sake of illustration, we have used a layered air-silver-air geometry with a silver slab of thickness $20nm$, for which we have fitted the Drude model (\ref{eq:drudeModel}) to the experimental data of silver within the $327-407 nm$ range, where $\w_p = 1.64\times 10^{16}\ rad/s$, $\gamma = 1.36 \times 10^{14}\ rad/s$, $\e_\infty = 7.45$ \cite{babar2015optical}. On one side of the slab, a point source (a vertical electric dipole) is positioned $20nm$ away from the lens, while the fields of the image are observed $20nm$ away from the interface on the other side.

\begin{figure}[htbp]
\centerline{\includegraphics[width=16cm]{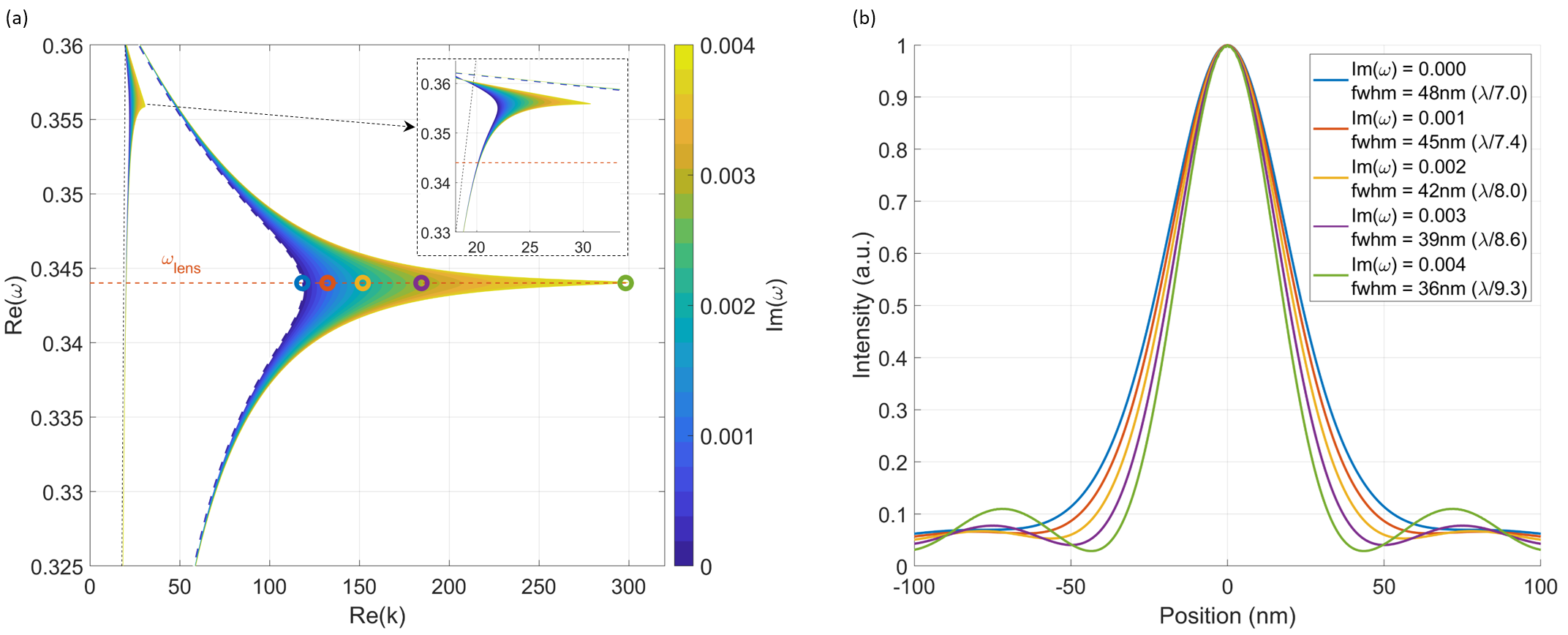}}
\caption{\textbf{(a)} The dispersion surface for the air-silver-air (IMI) structure, i.e. the superlens. Since the superlens supports two SPP modes, the dispersion surface consists of two distinct regions/surfaces, one for each mode. The higher frequency SPP (see inset) does not contain high spatial frequencies. The lower frequency SPP, however, extends to large $k$ values and its backbending region indeed defines the operating frequency ($\w_{lens}$) of the superlens where $Re(\e_{silver}) \approx -\e_0$. The colored circles indicate the complex frequencies with different exponential decay rates that are selected for the FDTD simulations. \textbf{(b)} Using the FDTD simulations, image intensities are measured for each frequency selected on the dispersion surface (colored circles), with their full-widths at half-maximum (FWHM). It is seen that the faster decay rates provide better resolution. As predicted by the dispersion surface, the SPP modes excited with larger $Im(\w)$ have larger wave vectors which, in turn, leads to a narrower point spread function.}
\label{superlens_dispersion_surface_and_fwhm}
\end{figure}

Following the same procedure we outlined for the MIM waveguide, we have calculated the dispersion surface for the superlens, as shown in Figure \ref{superlens_dispersion_surface_and_fwhm}a. Notice that the dispersion surface consists of two distinct regions corresponding to the two SPP modes supported by the insulator-metal-insulator (IMI) structure. The low frequency mode extends to very large values of $k$, making up the high spatial frequency components required for a high resolution image. As such, the backbending tip corresponds to the operating frequency of the superlens ($\w_{lens}$) where the permittivity of the silver matches the surrounding medium, i.e. $Re(\e_{silver}) \approx -\e_0$. The high frequency mode (see inset), on the other hand, bends back at significantly  lower values of $k$ than the operating frequency of the superlens, and therefore, does not play any role in the operation of the superlens.

To demonstrate the improvement in the resolution of the superlens due to the introduction of temporal loss into the system, we have selected five different exponential decay rates in time, as indicated by the circles on the dispersion surface in Figure \ref{superlens_dispersion_surface_and_fwhm}a, at the operating frequency of the superlens. Then, their image intensities are obtained from the FDTD simulations on a line parallel to the interface of the superlens, with their full-widths at half maximum (FWHM) calculated and shown in Figure \ref{superlens_dispersion_surface_and_fwhm}b. As predicted by the dispersion surface, faster decay rate in time provides significantly better resolution in space because the excited SPP mode attains a larger characteristic wave vector. In this example, we have been able to improve the resolution of the superlens for a point object by $25\%$, from $48nm$ to $36nm$. In conclusion, we have demonstrated that, with the help of the dispersion surface, we can improve the resolution of the superlens, and can easily apply the approach to similar problems like those studied in \cite{rogov2018space, dubois2015time, archambault2012superlens}.

\section{Discussion}

In this study, it has been demonstrated theoretically, as well as numerically, that the resolution and the propagation length of SPPs in plasmonic structures can be improved by introducing temporal loss into the system. Although this sentence captures the main highlights of the study, the underlying principle needs to be emphasized for the sake of clarity. The whole idea of introducing temporal loss into plasmonic layered media started from a long lasting confusion of how to decide on the permissible ranges of values of the frequency and the wavenumber in the solution set of the wave equation in such structures. However, in the mathematical study of wave equations with boundary conditions, it has been proven that the solutions in layered open geometries are in the form of $e^{j\w t-jkx}$ (also referred to as eigen-mode) where the permissible ranges of $\w$ and $k$ are over the entire complex plane unless the boundary conditions dictate otherwise. Contrary to what mathematics dictates on the choices of $\w$ and $k$ values, they have been generally assumed that either one has to be real, depending on the exact physical situation, excitation and experimental setup, and therefore, the solutions of the wave equation with simultaneous complex wave vectors and complex frequencies have been dismissed and not investigated so far. In this study, we have considered $e^{j\w t-jkx}$ with complex frequency $\w$ and complex wave vector $k$ as the fundamental mode of the plasmonic waveguide, and showed that they have properties superior to the SPPs with real frequencies or real wave vectors.
In addition, having the fundamental mode represented in exponential form with both $\w$ and $k$ complex enables us to utilize the conventional frequency-domain tools and results, such as the Fresnel reflection coefficient, the Drude model, and the dispersion equation, by just substituting in complex $\w$ and $k$. As a result, a simple yet powerful framework for pseudo-monochromatic modes has been established and verified using the FDTD simulations. The previous approaches that studied time-domain signals to improve the resolution of plasmonic devices required either time-domain solutions or numerically expensive frequency-domain integrals \cite{rogov2018space, dubois2015time, archambault2012superlens}.

\begin{figure}[htbp]
\centerline{\includegraphics[width=10cm]{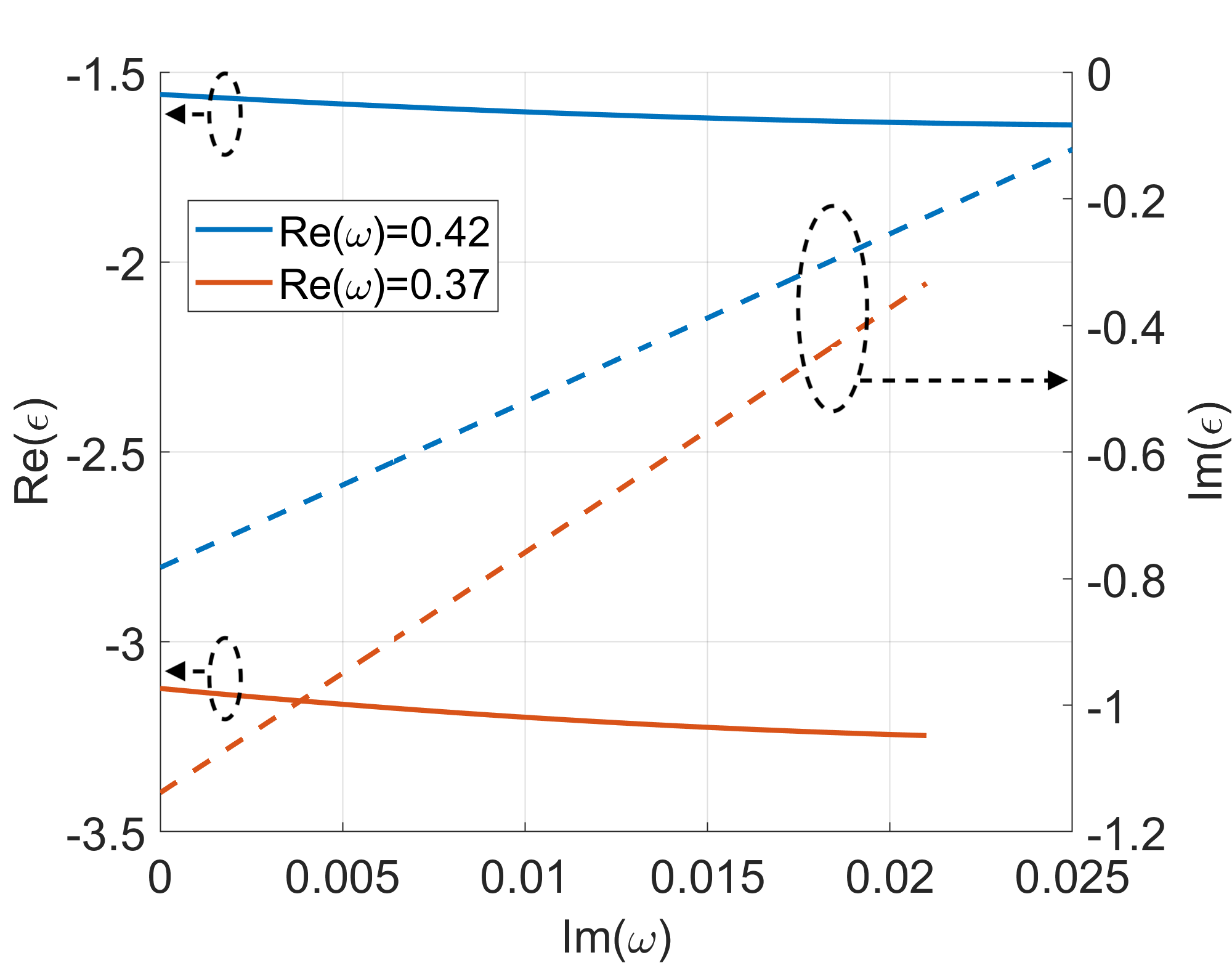}}
\caption{Relative permittivity of the metal by the Drude model for a range of complex frequency (\ref{eq:drudeModel}), where $\w_p = 1.04\times 10^{16}\ rad/s$, $\gamma = 6.15 \times 10^{14}\ rad/s = 0.059 \times \w_p$, $\e_\infty = 4$ are used. Note that $|Im(\e)|$ decreases and $|Re(\e)|$ increases as temporal decay $Im(\w)$ is introduced to the excitation source. This is the underlying reason for the improvements observed in the resolution and propagation length.}
\label{drude_model_complex_w}
\end{figure}

Another issue that needs further clarification is the counterintuitive improvements in resolution and propagation length of the SPPs when the source is turned off exponentially in time, in other words, when the temporal loss is introduced into the system. The underlying physics responsible for these phenomena is actually hidden in the equation of motion for free electrons, which is used to derive the Drude model for the relative permittivity of the metal, whose real and imaginary parts are shown in Figure \ref{drude_model_complex_w} for a range of complex frequencies. While the increase in the effective propagation length can be attributed to the decrease in the magnitude of the imaginary part of the permittivity (dashed lines), the resolution benefits from both the decrease in $|Im(\e)|$ and the increase in $|Re(\e)|$ (solid lines). We have found that there is a theoretical upper limit on resolution, which is exactly at the natural resonance frequency of the electrons, i.e., at $Im(\w)=\gamma/2$, where $Im(\e)$ is zero and $|Re(\e)|$ reaches its maximum. However, there is no such limit for the propagation length, except for many practical limits arising as the temporal decay rate increases.

Considering the results in Figure \ref{MIM_disp_surface_combined} and Figure \ref{MIM_resolution_curve}, it is possible to prescribe a recipe for manipulating the characteristics of an SPP. If a wide tuning range for the resolution is desired, the frequencies closer to (and below) the back-bending region are more advantageous. This comes at the cost of propagation length since the SPPs at those frequencies experience relatively higher loss compared to lower frequency SPPs. On the other hand, if one desires to maximize the propagation length, one should resort to the lower frequencies away from the back-bending region. This will provide a means to significantly increase the propagation length without going into the extreme $Im(\w)$ values. There will also be a slight increase in the resolution, although it is minor compared to the higher frequency SPPs.

Another advantage of the pseudo-monochromatic framework presented in this study is the simplicity of the source used to excite the complex SPP modes. Ultrafast pulse shaping techniques are already available for even more complex waveform generation \cite{ferdous2011spectral,weiner2011ultrafast}. Therefore, we do not expect a significant problem in realizing practical sources with complex frequencies. Moreover, fluorescent molecules might also provide a natural source of pseudo-monochromatic light excitation since their temporal decays are always exponential in nature. Finally, already existing experimental methods, such as the Kretschmann-Raether, Otto configurations or electron beams, or their modified versions may be employed to selectively control the frequency, wave vector, or the propagation speed of the SPPs \cite{maier2007plasmonics,gong2014electron}.


\section{Methods}

\subsection{Numerical simulation of plasmonic waveguides and identification of SPP modes}

The open waveguide (see Figure \ref{MIM_waveguide_dipole}) is simulated in a 2D FDTD region with the dimensions of $6um$ along the waveguide direction ($x$), and $4um$ in the vertical direction ($z$). The simulation region is surrounded by perfectly matched layers (PML) on all four sides in order to mimic the open structure with minimal artificial reflections from the PMLs. A uniform square mesh with a side length of $0.5nm$ covers the $50nm$ thick insulating layer (air in this case) and extends $175nm$ into the metal layers on both sides. For the rest of the metal layers, a conformal mesh that gradually becomes less dense away from the interfaces is used to reduce the computation time. With regard to the source and the resulting field distributions, a vertical electric dipole is placed in the insulating layer, while a time-domain line monitor is used in the same layer along the propagation direction to collect the field components as a function of both space and time.

\begin{figure}[htbp]
\centerline{\includegraphics[width=14cm]{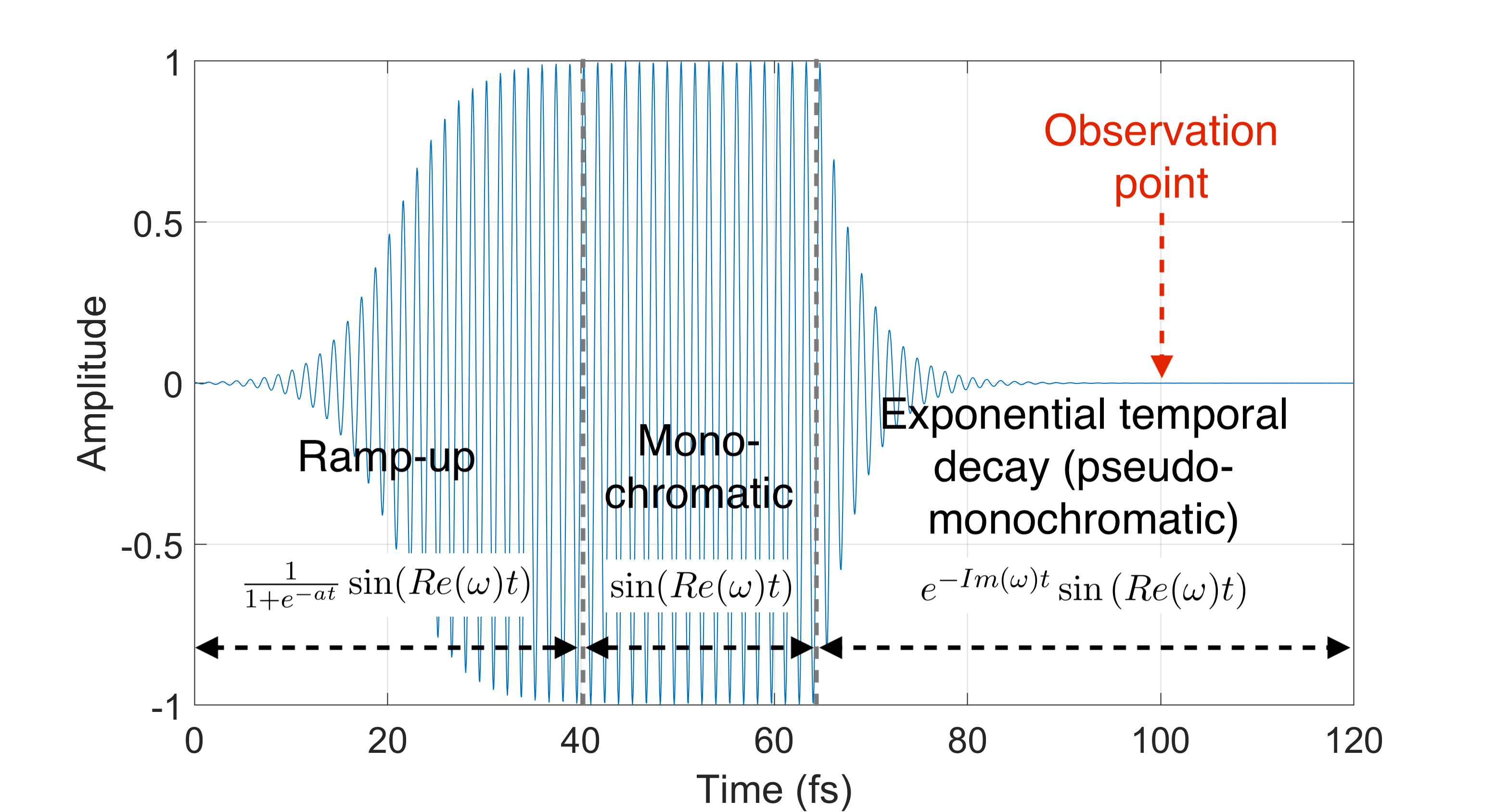}}
\caption{The dipole excitation signal used in the FDTD simulations. The oscillation amplitude is slowly increased in order to avoid undesired numerical noise in simulations. Then, the system is driven monochromatically until it reaches the steady-state, making sure the transient fields have decayed. Finally, the exponential temporal decay is applied to the dipole. The resulting complex excitation frequency generates a complex SPP mode with improved resolution and propagation length. The fields are observed long after the start of the pseudo-monochromatic signal, making sure any transient fields due to switching do not affect the measurements.}
\label{MIM_dipole_field}
\end{figure}

The waveform used to excite the waveguide is shown in Figure \ref{MIM_dipole_field}, where the time signature of the dipole source is shown to oscillate with a frequency of $Re(\w)$ at all times with an amplitude modulation. Initially, the amplitude is slowly increased from zero to a finite value, by modulating it with the sigmoid function $\frac{1}{1+e^{-at}}$. This is to alleviate possible convergence issues that may be encountered in the FDTD simulations when the source is turned on abruptly, resulting in an excitation of a wide spectrum of frequencies that might in turn give rise to unwanted reflections from the PML. Note that this slow buildup is needed only for the FDTD simulations and has no relevance to any real-life applications. Following that, the dipole is let to oscillate for a while at this constant amplitude so that the transient fields generated by the ramp-up attenuate significantly. When the steady-state is reached, an exponential temporal decay $e^{-Im(\w)t}$ is introduced in order for effectively making the oscillation frequency complex. As a result, a complex SPP mode with improved resolution and propagation length can be excited. To detect this SPP mode, the fields are observed long after the exponential decay is applied (indicated by the red arrow in Figure \ref{MIM_dipole_field}), making sure the transient fields due to switching from monochromatic to pseudo-monochromatic region have died off.

The field components of the SPP are collected on a $1\mu m$ long line along the waveguide away from the dipole. In order to extract the frequency and the wave vector of the SPP from the field distribution, we use the generalized pencil-of-function (GPOF) method, which helps represent the field in terms of complex exponentials~\cite{hua1989generalized}, whose exponents are proportional to $kx-\w t$. That is, the field distribution monitored along the $1\mu m$ line at a fixed time, which contains approximately 5 periods of the SPP, is decomposed into two complex exponentials, one for the SPP and the other for the numerical noise or any remaining lossy waves. Since the waveguide supports only the SPP mode at such a distance, two exponentials are sufficient to obtain accurate results of $k$. The same approach is also used to retrieve the complex $\w$ from the same field distribution, but this time at a fixed distance $x$ over a finite time span.

For the sake of demonstration, we ran a few numerical experiments following the outlined procedure for the source with the temporal decay of up to $Im(\w)=0.025$, as normalized to the plasma frequency $\w_p$, since faster decays require finer and larger FDTD simulations. We believe that the results presented in this study are sufficient to showcase this new venue for dispersion engineering, and to convey the underlying physics. As discussed in the main text, it is possible to obtain even further improvements than shown in Figure \ref{MIM_resolution_curve}~\&~\ref{MIM_effective_propagation_length} by employing faster temporal decay rates.




\bibliography{myRefs}{}
\bibliographystyle{naturemag}

\end{document}